# Activities of Women in Physics Group in Spain (2022–2023)

P. García-Martínez[1, a)], C. Ocal[2], A. X. López[3], M. Tórtola[4],
M. F. Morcillo-Arencibia[5], A. Martín-Molina[6], and S. Estradé[7]

[1] *Optics Department. Faculty of Physics. University of Valencia. Campus Burjasot-Paterna, 46100 Valencia, SPAIN.*
[2] *Materials Science Institute of Barcelona (ICMAB-CSIC), Campus UAB-Bellaterra 08190 Barcelona, SPAIN.*
[3] *Higher Polytechnic University College, University of A Coruña, 15471 Ferrol, SPAIN.*
[4] *Theoretical Physics Department and IFIC, CSIC-Universitat de València, 46980 Paterna. SPAIN.*
[5] *Physics Department. Faculty of Sciences. University of Cordoba. Campus Rabanales, 14071 Cordoba. SPAIN.*
[6] *Applied Physics Department, University of Granada, Campus Fuentenueva, 18071 Granada, SPAIN.*
[7] *Department of Electronic and Biomedical Engineering. University of Barcelona, 08028 Barcelona, SPAIN.*

Corresponding author: pascuala.garcia@uv.es

**Abstract.** In this paper, we present the main actions of the Women in Physics Group of the Spanish Royal Physics Society over the period of 2022–2023, in which we celebrated the 20$^{th}$ anniversary of the group. We also outline relevant equality initiatives implemented during this period by the Spanish Government as well as analyse their impact on the status of women in Physics in our country. In 2023, our scientific society approved the Gender Equality Plan, thus becoming a pioneer scientific society in Spain in implementing this relevant measure.

## SCIENTIFIC POLICIES TO PROMOTE WOMEN IN SCIENCE IN SPAIN

Two remarkable laws regarding scientific policy have been approved by the Spanish parliament in the period 2022–2023. One is the **Science, Technology and Innovation Law** (LCTI 17/2022)[1] that seeks to resolve the main gender inequalities that still persist in the field of research and development. Currently, all entities of the national public sector, from universities to research centers and funding agents, must have a Gender Equality Plan (GEP), including protocols against sexual harassment, and for reasons of sexual orientation, gender identity or sexual characteristics. The monitoring of GEPs will be carried out annually by the institutions. The inclusion of the gender dimension in projects is now mandatory as are the development measures to eliminate gender biases and guarantee equality in the selection and evaluation processes. Positive action measures will be promoted to support reconciliation of work and family life. Moreover, scientific dissemination and education are acknowledged as crucial in driving sociocultural change and fostering co-responsibility. The second law is the **Organic Law of the University System** (LOSU 2/2023)[2], which encourages the promotion of scientific projects with a gender perspective as well as gender parity in research teams and mechanisms facilitating a higher number of women principal investigators. In this context, according to the Gender Equality Index (GEI)[3] from the European Institute of Gender Equality, Spain reaches 76.4 points out of 100, which places it 4$^{th}$ in the European Union in GEI and 6.2 points above the score for the EU27 as a whole.

## NUMBERS OF WOMEN IN PHYSICS IN SPAIN

The Women in Physics Group (GEMF) of the Spanish Royal Physics Society (RSEF) was created in 2002, following the IUPAP Working Group of Women in Physics guidelines in the Resolution of the 23$^{rd}$ General Assembly of IUPAP, Atlanta, Georgia, (1999). Currently, 21 % of RSEF members are women and the GEMF is made up of 70 % women and 30 % men who work together with the objective of increasing the presence of women in Physics in the Spanish context, to make their achievements visible, to report any situation of inequality and to firmly defend the interests, equal rights and opportunities of women in Physics. In this regard, it is important to highlight that in 2023, the RSEF approved its first GEP[4], becoming a pioneer scientific society in Spain. In this GEP, the RSEF positions itself to give visibility and support to women in Physics, through five lines of action with specific measures that commit all the areas and activities carried out within the RSEF.

---

[1] Ley de Ciencia, Tecnología e Innovación, LCTI 17/2022 https://www.boe.es/buscar/act.php?id=BOE-A-2023-7500/
[2] Ley Orgánica del Sistema Universitario, LOSU 2/2023 https://www.boe.es/buscar/act.php?id=BOE-A-2023-4513/
[3] Gender Equality Index https://eige.europa.eu/gender-equality-index/2023/country/ES/

The first action of the GEP is "To carry out a gender diagnostic of the RSEF". This study, which is currently underway, should be complemented by a more comprehensive analysis on the situation of women physicists in the professional as well as academic fields, in order to promote the participation of girls and women in conditions of full equality with their male colleagues. In this sense, the GEMF has produced a first report on the presence of women physicists in academia [1], and some results are shown in Fig. 1 together with those obtained in a study carried out by the Women and Science Commission of the National Research Council, the largest public research institution in Spain [2]. As can be seen from Fig. 1(a), the number of physics students in Spanish faculties varies over time, with the proportion of women enrolled fluctuating between 24 and 36 % over the period 1985–2021 (see Fig. 1(b)), although it has not exceeded 30 % in the last decade.. The data also show that the gap between the number of male and female graduates is widening.

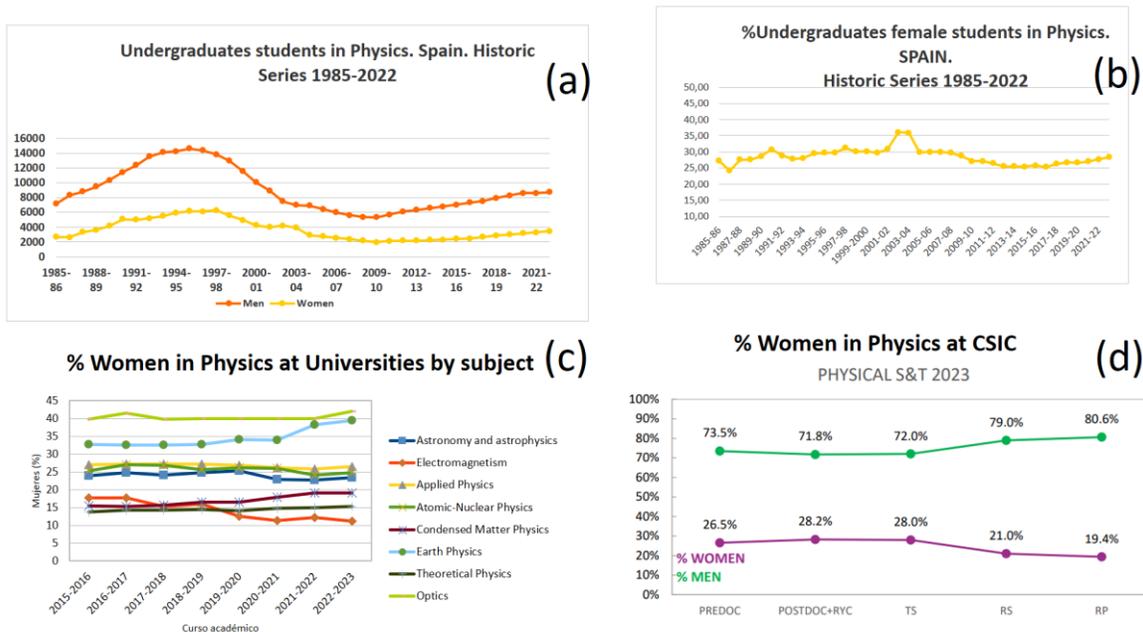

**FIGURE 1.** Number of (a) undergraduates and (b) percentages of female students in Physics in Spain (1985–2023). Percentages of women in Physics working at (a) Spanish universities by area and (d) National Research Council (CSIC) in 2023. Data Sources by CSIC Commission for Women in Science [2] and GEMF.

One possible argument for the shortage of women in Physics could be the underrepresentation and low visibility of female physicists in media and in popular culture. The lack of role models, coupled with the stereotype that physics is a male-dominated field, can discourage young girls from pursuing a career in physics. Unfortunately, the culture of physics and STEM fields in general can be quite unwelcoming to women, so that those who have chosen to pursue an academic career are more likely to experience discrimination, harassment, and bias than their male counterparts, as various studies have shown. This hostile environment can make it difficult for women to thrive in physics. In fact, the number of women in physics departments at Spanish universities is 25 % of the total teaching and research staff, remaining practically constant over the years (Fig. 1(c)). Likewise, the average is 22.2 % considering all research staff levels (Fig. 1(d)) in the Physical Science and Technology (S&T) area of the CSIC.

## ACTIVITIES AND ACTIONS

In 2022, the GEMF marked its 20th anniversary, highlighting its extensive history as a working group and as an active participant in the ICWIP conferences. The commemoration took place on December 16, 2022, at the Institute of Optics "Daza de Valdés". The program (Fig. 2(a)) included the participation of worldwide recognized women physicists. All sessions were recorded and can be accessed through our YouTube channel and our website[5]. The same year, the GEMF organized various activities as a part of the 38th Physics Biennial of the RSEF, which included a round table discussion on "*Advancing Gender Equality in Physics in Europe*," featuring the EU project GENERA and the Project JUNO from the UK's Institute of Physics. The main activity was the Women in Physics Symposium (Fig. 2(b)). Like every year, we organized webinars on topics concerning women in science, research, and education (Fig. 2(c)).

---

[4] Plan de Igualdad de la RSEF (2023) https://rsef.es/images/Fisica/PlanigualdadRSEF.pdf

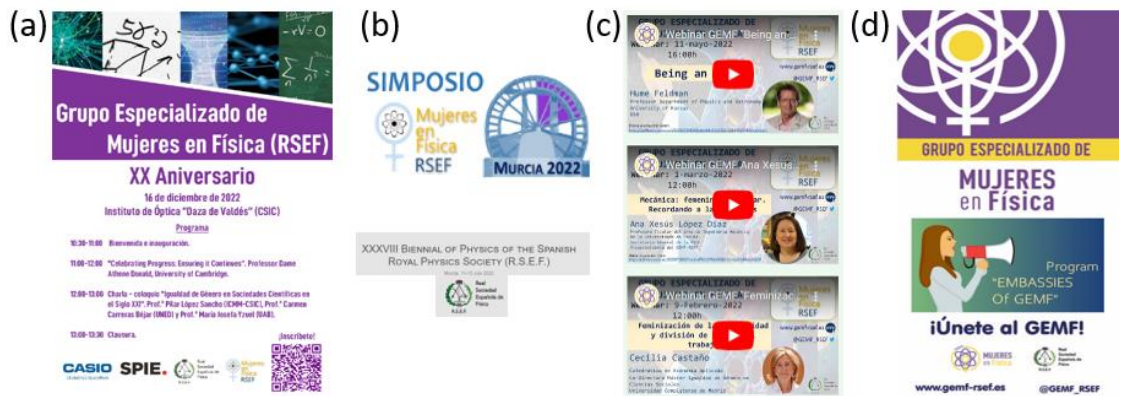

**FIGURE 2.** Period 2022–2023: (a) 20th anniversary celebration poster and program. (b) Announcement of the GEMF Symposium at the 38th Physics Biennial of the RSEF. (c) Some GEMF webinars. (d) Embassies of GEMF poster.

In 2023, as a commitment to the higher education sector, we launched the project "*Embassies of GEMF*" (Fig. 2(d)) with the involvement of GEMF volunteers at Spanish universities offering Physics degrees. This initiative aims to facilitate collaboration among physics departments at universities to enhance communication, training, and outreach activities related to gender and physics. Furthermore, the project provides a platform to establish local connections for sharing tools and resources to combat sexist stereotypes and biases within the field of physics. The Women in Physics Group in Spain has an active presence on social media, with a role committed to the promotion of women in science on Twitter/X (@GEMF_RSEF) and also as a tool to disseminate calls and report situations of bias in science (e.g., all-men panels in congresses, nonparity juries, etc.). The GEMF official website contains valuable information (from news to document repository), and all GEMF members receive at least one informative email per month with the latest worldwide news concerning women in science, highlighting specific reading and providing links of interest, thus keeping our current membership community updated.

---

[5] https://www.gemf-rsef.es/